\documentclass[prd,preprint,eqsecnum,nofootinbib,amsmath,amssymb,
               tightenlines,dvips]{revtex4}
\usepackage{graphicx}
\usepackage{bm}

\begin {document}


\def\ca{C_{\rm A}}

\def\cf{C_{\rm F}}

\def\Nf{N_{\rm f}}
\def\Nc{N_{\rm c}}
\def\ta{t_{\rm A}}
\def\tf{t_{\rm F}}
\def\md{m_{\rm D}}
\def\alphas{\alpha_{\rm s}}

\def\q{{\bm q}}

\def\gammaE{\gamma_{\rm E}^{}}

\def\tr{\operatorname{tr}}

\def\qgq{{q{\to}gq}}
\def\ggg{{g{\to}gg}}
\def\gqq{{g{\to}q\bar q}}
\def\abc{{a{\to}bc}}
\def\ellgE{\ell_{\rm g}^{\rm (e)}}
\def\ellqE{\ell_{\rm q}^{\rm (e)}}
\def\AgE{A_{\rm g}^{\rm (e)}}
\def\AqE{A_{\rm q}^{\rm (e)}}
\def\qhat{\hat{\bar q}}
\def\hatmd{\hat m_{\rm D}}
\def\hatN{\hat{\cal N}}


\title
    {
      Stopping distance for high energy jets in weakly-coupled
      quark-gluon plasmas
    }

\author{
  Peter Arnold,
  Sean Cantrell,%
   \footnote{
     Current address:
     Department of Physics \& Astronomy,
     Johns Hopkins University,
     3400 N. Charles St.,
     Baltimore, MD 21218
   }
  and Wei Xiao
}
\affiliation
    {%
    Department of Physics,
    University of Virginia, Box 400714,
    Charlottesville, Virginia 22904, USA
    }%

\date {\today}

\begin {abstract}%
{%
   We derive a simple formula for the stopping distance for a
   high-energy quark traveling through a weakly-coupled quark
   gluon plasma.  The result is given to next-to-leading-order
   in an expansion in inverse logarithms $\ln(E/T)$, where $T$
   is the temperature of the plasma.
   We also define a
   stopping distance for gluons and give a leading-log result.
   Discussion of stopping
   distance has a theoretical advantage over discussion of energy loss rates
   in that stopping distances can be generalized to the case of
   strong coupling, where one may not speak of individual partons.
}%
\end {abstract}

\maketitle
\thispagestyle {empty}


\section {Introduction and Results}
\label{sec:intro}

There has long been interest in calculating how a fast moving quark or
gluon looses energy when traveling through a quark-gluon plasma.
One of the simpler versions of this problem is to calculate energy
loss in the case of a spatially-infinite equilibrium plasma
(i.e.\ a plasma whose size is large compared to the stopping distance).
For weakly-coupled plasmas, energy loss is dominated by hard bremsstrahlung
and pair production in the high-energy limit $E \gg T$.
Results for the rates of these processes at leading order in $\alphas$ have
been known for some time, as we'll review in a moment.
The rate is parametrically \cite{BDMPS1,BDMPS2,BDMPS3,Zakharov1,Zakharov2}
\begin {equation}
   \Gamma \sim \alphas^2 T \sqrt{\frac{\ln(E/T)}{E/T}}
\label {eq:brem0}
\end {equation}
for hard bremsstrahlung and pair production at high energy
(with some caveats on the range of applicability to be discussed
later).
The roughly
$E^{-1/2}$ decrease of this rate with energy is a result of the
Landau-Pomeranchuk-Migdal (LPM) effect \cite{LP,Migdal},
which arises from
interference between bremsstrahlung
(or pair production) associated with multiple collisions of the
particle with the plasma.  If there were no LPM effect, then the
high-energy rate would have been approximately independent of $E$.

From the rate of bremsstrahlung and pair production, one can
figure out how far a high-energy particle travels before it ``stops'' in
the plasma, by which we mean that the particle loses
enough energy to reach $E \sim T$ and so equilibrate with the plasma.
Consider the energy loss of a quark.  Crudely, if the quark loses
roughly half its energy in the first hard bremsstrahlung, and roughly
half in the next, and so forth, then the roughly $E^{-1/2}$
behavior of the rate (\ref{eq:brem0})
suggests that the stopping distance
is parametrically of order
\begin {align}
   \ell_{\rm stop}
   &\sim \frac{1}{\Gamma(E)} + \frac{1}{\Gamma(E/2)}
        + \frac{1}{\Gamma(E/4)}
        + \frac{1}{\Gamma(E/8)} + \cdots
\nonumber\\
   &\sim \frac{1}{\Gamma(E)} \left( 
        1 + \frac{1}{2^{1/2}} + \frac{1}{2}+ \frac{1}{2^{3/2}} + \cdots
    \right)
\nonumber\\
   &\sim \frac{1}{\Gamma(E)}
   \sim \frac{1}{\alphas^2 T} \sqrt{\frac{E/T}{\ln(E/T)}} \,.
\label {eq:t0}
\end {align}

In weak coupling, the $E^{-1/2}$ fall-off of the rate is due to the
LPM effect.  It's interesting to ascertain whether anything similar happens
in plasmas that are not weakly coupled.
Fortunately, there are
certain strongly-coupled gauge theories where people have,
using AdS/CFT duality, calculated
results related to the slowing down of high-momentum particles.
We cannot directly discuss bremsstrahlung or pair production rates
at strong coupling,
however, because bremsstrahlung and pair production implicitly refer
to individual quanta and so are intrinsically perturbative concepts.
But one can generalize the idea of stopping distances to non-perturbative
situations, as nicely explained by
Chesler, Jensen, Karch, and Yaffe \cite{Chesler},
as we
shall review later.  In the strongly-coupled gauge theories that have
been studied, stopping distances at high energy behave like $E^{1/3}$
\cite{Chesler,Gubser,Hatta} rather than $E^{1/2} / \sqrt{\ln E}$.

Because stopping distances, unlike bremsstrahlung rates, can be generalized
to strongly-coupled situations, it's interesting to know how to
calculate them in weakly-coupled situations.
In this paper, we derive simple results for various stopping distances in
QCD in the high-energy limit $\ln(E/T) \gg 1$, working to leading order
in coupling $\alphas$.  Though we assume $\ln(E/T) \gg 1$, we will
formally assume that $\alphas$ is small enough that
$\alphas \ln(E/T) \ll 1$ and so, for instance, we ignore the
running of the coupling $\alphas$.
(We will say a few words later about how to plausibly modify the
answer to include running coupling effects.)

Our most thorough calculation is of the stopping distance for a quark,
which we compute to next-to-leading-logarithmic order (NLLO) in
$\ln(E/T)$.
Our results can be summarized as
\begin {subequations}
\label {eq:qstop}
\begin {equation}
  \ell_{\rm stop,q} \simeq \frac{1}{a \alphas^2 T} \sqrt{ \frac{E}{TL} } ,
\label {eq:qstop1}
\end {equation}
where the logarithm $L$ is determined by self-consistent solution of
the equation
\begin {equation}
  L = \ln\left(\frac{b E L}{\alphas^c T}\right)
\label {eq:L}
\end {equation}
\end {subequations}
and $a$, $b$, and $c$ are numerical constants.
These constants are given in Table
\ref{tab:abc} in the large $E$ limit of
$E \gg T/\alphas^2 \ln(\alphas^{-1})$,
or in Table \ref{tab:abcsmall} in the case
$T \ll E \ll T/\alphas^2 \ln(\alphas^{-1})$ of smaller
(but still formally large) $E$.
The latter case has $c=0$ in (\ref{eq:L}) and so corresponds to
the simple parametric estimate (\ref{eq:t0}) based on (\ref{eq:brem0}).

\begin {table}[t]
\begin {tabular}{|c c||c|c|c||c|c|}
\hline
   & $\Nf$ & $a$ & $b$ & $c$ & $a^{\rm(e)}_{\rm g}$ & $a^{\rm(e)}_{\rm q}$ \\
\hline
QCD
 & 0 & --- 
     & --- 
     & 0.736866 & 5.26606 & ---
     \\
 & 2 & 2.06480 & 0.092141 & 0.432769   & 5.98760 & 3.61355 \\
 & 3 & 2.23024 & 0.105282 & 0.345885   & 6.27235 & 3.87544 \\
 & 4 & 2.38423 & 0.115487 & 0.280721   & 6.52525 & 4.11632 \\
 & 5 & 2.52886 & 0.123314 & 0.230039   & 6.75427 & 4.34031 \\
 & 6 & 2.66565 & 0.129246 & 0.189492   & 6.96480 & 4.55038 \\
 \hline
QED
 & 1 & 0.30216 & 3.16393  & -0.175423  & 0.11121 & 0.12072 \\
\hline
\end {tabular}
\caption {
  \label {tab:abc}
  Coefficients $a$,$b$,$c$
  for the quark-number stopping distance (\ref{eq:qstop})
  in the case $Q_\perp \gg T$ (parametrically
  $E \gg T^4/\hat q$ or equivalently
  $ E \gg T/\alphas^2 \ln(\alphas^{-1})$).  Also shown are
  the leading-log coefficients
  $a^{\rm(e)}_{\rm g}$ and $a^{\rm(e)}_{\rm q}$  for the
  gluon and quark energy stopping distances defined in
  Sec.\ \ref{sec:energydef}.  Note that larger values of $a$
  correspond to shorter stopping distances in the high-energy limit.
}
\end {table}

\begin {table}[t]
\begin {tabular}{|c c||c|c|c||c|c|}
\hline
   & $\Nf$ & $a$ & $b$ & $c$ & $a^{\rm(e)}_{\rm g}$ & $a^{\rm(e)}_{\rm q}$ \\
\hline
QCD
 & 0 & --- 
     & --- 
     & 0   & 6.16024 & ---
     \\
 & 2 & 2.27727 & 0.533221 & 0   & 6.60372 & 3.98538 \\
 & 3 & 2.41541 & 0.473974 & 0   & 6.79311 & 4.19719 \\
 & 4 & 2.54606 & 0.426577 & 0   & 6.96816 & 4.39572 \\
 & 5 & 2.67033 & 0.387797 & 0   & 7.13213 & 4.58313 \\
 & 6 & 2.78907 & 0.355480 & 0   & 7.28728 & 4.76106 \\
\hline
QED
 & 1 & 0.28861 & 2.92883  & 0   & 0.10622 & 0.11531 \\
\hline
\end {tabular}
\caption {
  \label {tab:abcsmall}
  Coefficients for the stopping distance
  in the case $Q_\perp \ll T$ (parametrically
  $T \ll E \ll T^4/\hat q \sim T/\alphas^2 \ln(\alphas^{-1})$).
}
\end {table}

Later, we will also give a definition of a stopping distance for gluons
that can be generalized to non-perturbative gauge theories, and we will
give a result to leading-log order for that distance in the case of
weak coupling.

At the moment, we do not have a precise weakly-coupled result for
the types of supersymmetric gauge theories in which the strongly-coupled
limit has been studied.  The answer will be in the form (\ref{eq:qstop})
for weak coupling, but we leave the calculation of the coefficients
$a$, $b$, and $c$ in this case for future work.

Throughout this paper, we will assume that the baryon chemical potential
of the quark-gluon plasma is negligible compared to its temperature.

In section \ref{sec:define}, we review
Chesler et al.'s definition of a stopping distance for quarks,
which we call the quark number stopping distance, and then we discuss
a natural generalization for discussing gluons, which introduces what
we will call the quark and gluon energy stopping distances.
Section \ref{sec:LL} derives results for these various stopping
distances to leading-log order, which is enough to determine
the coefficients $a$ and $c$ in our result (\ref{eq:qstop}).
Section \ref{sec:NLLO} then
continues to next-to-leading log order for the quark number
stopping distance, giving the remaining coefficient $b$.
As a test of our NLLO result, we compare it in
section \ref{sec:numerics} to
numerical simulations of energy loss that make no approximation
regarding the size of $\ln(E/T)$ but instead use full
weak-coupling results for the bremsstrahlung rate.  We verify that
our NLLO result (\ref{eq:qstop})
correctly reproduces the large $E$ behavior.
Finally, in section \ref{sec:running}, we discuss how one might
modify our result to account for running of $\alphas$.


\section{Defining Stopping Distances}
\label {sec:define}

\subsection{Quark Number Stopping}

\begin {figure}
\begin {center}
  \includegraphics[scale=0.6]{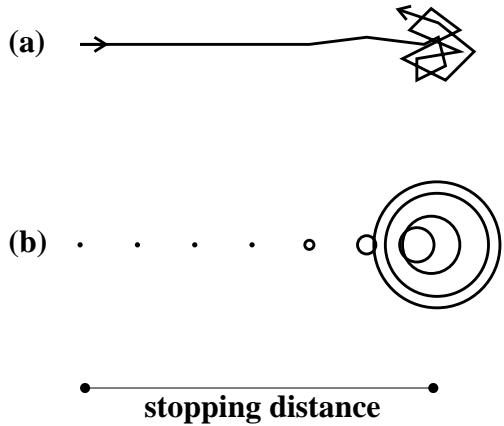}
  \caption{
     \label{fig:path}
     The stopping of a high-energy quark represented by (a)
     the path of an individual quark, and (b) the motion of
     the center of the spreading quark number probability distribution.
     In (b), the quark number probability distribution at each
     successive time
     is represented by a {\it single}\/ contour.  As time progresses,
     the center of the contour moves from left to right, slowing down
     with time, and the
     diameter of the contour increases.
  }
\end {center}
\end {figure}

To explain Chesler et al.'s definition of a stopping distance
for quarks, we first focus on the case of weak coupling, where we
can talk about individual particles.
We gave a crude estimate of the stopping distance in
(\ref{eq:t0}), but we should consider that
an individual quark doesn't really stop in a weakly-coupled
plasma.  Once its energy drops
to $E \sim T$, then the quark is just as likely to gain energy from its
interactions with the plasma as lose energy.  A cartoon of an individual
high-energy quark's trajectory is shown in Fig.\ \ref{fig:path}a.
At first, the quark moves in a nearly straight line as it loses
energy.  But once its energy falls to be $\sim T$, it then
random walks through the plasma just like any other equilibrated
quark.  Suppose we repeat this thought experiment over and over again
in order to determine the probability {\it distribution}\/ of the
quark's position as a function of time. A cartoon of the time evolution
of this distribution is shown in Fig.\ \ref{fig:path}b.  Once the quark
equilibrates, the center of the distribution stops moving, and the
distribution merely spreads in size due to diffusion.  We can then
define the stopping distance of a quark as the distance that the
center of the probability distribution moves between $t{=}0$ and $t{=}\infty$.

So far, we have been speaking perturbatively about the position of
the quark.  The probability distribution of quark positions
easily generalizes to the non-perturbative case: simply refer instead
to the probability distribution for total quark number in the system,
since the quark number operator is defined non-perturbatively.
But so far we've specified our initial state perturbatively when
we said it was a single high-energy quark moving through the quark-gluon
plasma.  Chesler et al.\ realized that this can also be generalized
to the non-perturbative case.  Consider all possible initial states
that (i) differ from thermal equilibrium in a localized region of space,
(ii) contain total energy $E$ in that region in excess of the
thermal background, and (iii) contain total
quark number one in that region.  Measure the stopping distance for that
initial state.  Then define the ``quark number stopping distance'' as the
maximum stopping distance, taken over all such choices of initial state.

Let's understand how this definition would work out in the
weakly-coupled case.  In our earlier discussion, our initial state
was a single quark carrying the total energy $E$.  What if we had
chosen a different initial state with the same energy and quark
number?  Consider, for instance, an initial state composed of a
quark of energy $E/2$ plus a gluon of energy $E/2$.  The quark
number stopping distance doesn't care about what happens to the gluon
and would just depend on how long it took the energy $E/2$ quark
to stop.  But according to (\ref{eq:t0}),
that is roughly $1/\sqrt2$ shorter time than if the
entire energy $E$ had been carried by a single quark.
So the initial state that gives the {\it maximum}\/ stopping distance
is indeed the one we want in the weakly-coupled case: all the energy
carried by a single quark.

The definition of stopping time just given implicitly assumed that
both the position and energy of the initial state are well defined.
Due to the uncertainty principle, however, the location of the initial
position will naturally be ambiguous by an amount of order $1/E$.
So the definition of stopping distance implicitly assumes the limit
that $\ell_{\rm stop} \gg 1/E$.

In the case of ${\cal N}{=}4$ large-$\Nc$ supersymmetric
Yang Mills theory with a massless ${\cal N}=2$
fundamental-charge ``quark'' hypermultiplet,
Chesler et al.\ find (with their non-perturbative definition of the
stopping distance)
\begin {equation}
   \ell_{\rm stop,q} =
   \frac{\cal C}{T} \left( \frac{E}{T\sqrt{\lambda}} \right)^{1/3}
\end {equation}
in the $\lambda \equiv \Nc g_{\rm s}^2 \to \infty$ limit,
with
\begin {equation}
  {\cal C} \approx 0.5 \,.
\end {equation}


\subsection{Gluon Energy Stopping}
\label {sec:energydef}

Even in weak coupling, there is some difficulty in defining exactly
what one means by the stopping distance of a gluon.  For quarks, it's
always clear what the energy of the quark is after a $q \to q g$
splitting process.  For gluons, we can have $g \to gg$ or
$g \to q\bar q$, and we have to decide which particle to follow.
One possibility would be to follow, at each splitting, the
final particle that has the highest
energy.  This definition does not have any obvious generalization
to strong coupling.  But we can instead use a definition like the
previous one for quark number stopping distance.  For the initial state,
we now require quark number zero.  Then, for the stopping distance, we
measure the distance traveled before
the center of the probability distribution for {\it energy}\/
in excess of equilibrium (rather than quark number) stops.
We will call this the ``gluon energy stopping distance.''

For a gluon in a weakly-coupled plasma, this definition
can be translated as follows.  Consider the splitting of an initial
high-energy gluon,  moving in the $z$ direction, which cascades through
splitting
into $N$ particles, which are ``stopped'' at
positions $z_1$, $z_2$, ..., $z_N$ relative to the initial position
of the initial gluon.  Here stopped means
that the individual energies $E_i$ are of order $T$.  The gluon
energy stopping distance is the energy-weighted average of these
stopping positions:
\begin {equation}
   \ell_{\rm stop, g}^{\rm (energy)}
   \simeq \frac{\sum_i E_i z_i}{E} \,.
\label {eq:tgdef}
\end {equation}
This simple translation is fuzzy when it comes to deciding exactly when
a final parton has low enough energy to be considered ``stopped.''  Is
it $E_i \le T$ or $E_i \le 2 T$ or ...?  However, because the time
scale $(\alpha^2 T)^{-1}$ for particles with energy of order $T$ to
significantly
interact is so much faster than the time scale (\ref{eq:t0}) for
the initial gluon to lose significant energy, this ambiguity only
involves corrections to the calculation of the stopping distance
that are suppressed by a factor of $\sqrt{T/E}$ (up to logarithms).
We can ignore these corrections if we are just interested in $E \gg T$.
Note that, when $E_i \sim T$, we should also be thinking about
collisional energy loss, which is then competitive with bremsstrahlung
and pair production.  But the effects of collisional loss on the
stopping distance will similarly be suppressed by a factor of
$\sqrt{T/E}$.


\section{Leading Log}
\label {sec:LL}

\subsection {Quark Number Stopping: General Analysis}

We'll begin with a leading-log calculation of the quark
number stopping distance, which is fairly simple.
To motivate the approximation, consider the crude, heuristic
estimate made in (\ref{eq:t0}).  At leading-log order, we can
ignore the difference between a $\ln(E/T)$ appearing in
$\Gamma(E)$, a $\ln(E/2T)$ appearing in $\Gamma(E/2)$, and
a $\ln(E/4T)$ appearing in $\Gamma(E/4)$.  Since the series in
(\ref{eq:t0}) converges sufficiently rapidly (i.e.\ geometrically),
it is sufficient at leading-log order to replace the
logarithm in the bremsstrahlung rate (\ref{eq:brem0}) by
a constant:
\begin {equation}
   \Gamma(E) \sim \alphas^2 T \sqrt{\frac{\ln(E_0/T)}{E/T}}
\end {equation}
where $E_0$ is the initial energy of the quark.
In this approximation, the $E$ dependence of $\Gamma(E)$ is
simply $E^{-1/2}$, without any logarithm.
That simplification will be the key to a simple solution
for the stopping distance in what follows.

In more detail, the bremsstrahlung rate is a function of the momentum
fraction $x = \omega/E$ of the emitted gluon of frequency $\omega$.
(For hard bremsstrahlung in the large $E$ limit,
we can ignore the thermal mass of the gluon.)  In the
leading-log approximation, we can write
\begin {equation}
   \frac{d\Gamma(E)}{dx} \simeq
   \left(\frac{E_0}{E}\right)^{1/2}\frac{d\Gamma(E_0)}{dx}
   \equiv E^{-1/2} \, \frac{d\tilde\Gamma}{dx} \,,
\label {eq:LLscaling}
\end {equation}
where there is an implicit logarithmic dependence of $d\tilde\Gamma/dx$
on $E_0$.
We will review the specific leading-log formula for
$q\to gq$ bremsstrahlung later.  But for now we can proceed
with the analysis in a general way, based simply on the scaling
(\ref{eq:LLscaling}).

Let $\ell_{\rm q}(E)$ be the quark number stopping distance as a function
of $E$.  We can formally write the following self-consistent
equation for $\ell_{\rm q}$:
\begin {equation}
   \ell_{\rm q}(E) =
   \frac{1}{\Gamma_\qgq(E)}
   + \int_0^1 dx \> \frac{ d\Gamma_\qgq(E,x)/dx }{ \Gamma_\qgq(E) }
     \> \ell_{\rm q}\bigl((1-x)E\bigr) .
\label {eq:tq}
\end {equation}
The first term on the right-hand side is the average distance before
the first bremsstrahlung.  That first bremsstrahlung splits the
quark into a nearly-collinear gluon with energy $xE$ and quark with
energy $(1-x)E$.  The second term on the right-hand side of
(\ref{eq:tq}) is the remainder of the stopping distance after the
first bremsstrahlung.  This is just $\ell_{\rm q}((1-x)E)$ weighted
by the probability $(d\Gamma/dx)/\Gamma$ that the first bremsstrahlung
had gluon momentum fraction $x$.  To find the stopping distance, we
just need to solve this equation in leading-log approximation.

Note that (\ref{eq:tq}) ignores what
happens to the radiated gluons.  Suppose a bremsstrahlung gluon later
splits into a $q\bar q$ pair with the new quark carrying more energy
than the new anti-quark.  The subsequent cascading of that pair will
then affect the distribution of quark charge at the end of the
cascade process, and so will affect where the center of the total
quark number distribution stops.  However, it's just as likely that
the gluon would have instead split into a $q\bar q$ pair in which the new
anti-quark carries more energy, and this will produce an opposite
effect on the location of the center of the final quark number
distribution.  Because of charge symmetry, we can ignore what
happens to the gluons when calculating the quark number stopping
distance.  We dwell on this point because the situation will
be different when we later compute the gluon energy stopping distance.

The reason we referred to (\ref{eq:tq}) as ``formal'' is that the
total bremsstrahlung rate
\begin {equation}
  \Gamma_\qgq(E) = \int_0^1 dx \> \frac{ d\Gamma_\qgq(E,x) } { dx }
\end {equation}
is infrared ($x{\to}0$) divergent in leading-log approximation.
You can imagine temporarily imposing some sort of infrared regulator.
But it is easy to manipulate (\ref{eq:tq}) to remove the need for
such a regulator by multiplying both sides by $\Gamma_\qgq(E)$ and
rearranging terms to rewrite the equation as
\begin {equation}
   \int_0^1 dx \> \frac{d\Gamma_\qgq(E,x)}{dx}
   \left[ \ell_{\rm q}(E) - \ell_{\rm q}\bigl((1-x)E\bigr) \right]
   =
   1 .
\label {eq:tq2}
\end {equation}
The fact that the factor in brackets vanishes as $x \to 0$ turns out
to be sufficient to make the $x$ integration in (\ref{eq:tq2})
infrared convergent.

Now make the leading-log approximation by substituting
(\ref{eq:LLscaling}) into (\ref{eq:tq2}) to get
\begin {equation}
   E^{-1/2} \int_0^1 dx \> \frac{d\tilde\Gamma_\qgq(x)}{dx}
   \left[ \ell_{\rm q}(E) - \ell_{\rm q}\bigl((1-x)E\bigr) \right]
   =
   1 .
\label {eq:tq3}
\end {equation}
We can solve this equation by writing
\begin {equation}
  \ell_{\rm q}(E) = \frac{E^{1/2}}{A_{\rm q}} ,
\end {equation}
where $A_{\rm q}$ is a constant.  Then the factors of $E$
drop out of (\ref{eq:tq3}), leaving
\begin {equation}
   A_{\rm q} =
     \int_0^1 dx \> \frac{d\tilde\Gamma_\qgq(x)}{dx}
     \left[ 1 - (1-x)^{1/2} \right] .
\label {eq:LLAq}
\end {equation}
It only remains to plug in the detailed leading-log formula for
$d\tilde\Gamma_\qgq/dx$ in leading-log approximation.  We'll leave that
to Sec. \ref{sec:LLdetails}.


\subsection {Gluon Energy Stopping: General Analysis}

For the gluon energy stopping distance $\ellgE(E)$
given in weak coupling (and high energy) by (\ref{eq:tgdef}),
we can write down an equation similar to (\ref{eq:tq}).
One difference is that we now have to account for both
of the high-energy particles present after the first splitting,
because they both carry energy.  To warm up, first consider
pure Yang Mills gauge theory with no quarks.  Then the only relevant
splitting process would be $g \to gg$, and the equation for
the gluon energy stopping distance would be
\begin {equation}
   \ellgE(E) =
   \frac{1}{\Gamma_\ggg(E)}
   + \frac12 \int_0^1 dx \> \frac{ d\Gamma_\ggg(E,x)/dx }{ \Gamma_\ggg(E) }
     \left[
        x \, \ellgE(xE)
      + (1-x) \, \ellgE\bigl((1-x)E\bigr)
     \right] .
\label {eq:tgpure}
\end {equation}
The new feature to this equation is that the energy stopping distances
$\ellgE(xE)$ and $\ellgE\bigl((1-x)E\bigr)$ of the two gluons after
the first splitting are averaged, weighted by their energies, as
required by (\ref{eq:tgdef}).  (An energy-weighted average
of the energy stopping distances of
these two gluons is the same thing as an energy-weighted average
of the positions of all their stopped descendants.)
The overall factor of $\frac12$ in front of the integral in
(\ref{eq:tgpure}) is to avoid double counting states of
the final two, identical gluons.

Once quarks are introduced, there is the added complication that the
energy carried by a gluon could be converted into energy carried by
quarks via $g\to q\bar q$.  To find the gluon energy stopping distance
$\ellgE(E)$, we will therefore need to also introduce the
quark energy stopping distance $\ellqE(E)$, which is given in
weak coupling by (\ref{eq:tgdef}) but in the case where the initial
particle is a quark (or anti-quark).
The basic equation (\ref{eq:tgpure}) for $\ellgE$ now becomes
\begin {multline}
   \ellgE(E) =
   \frac{1}{\Gamma_{\rm g}(E)}
   + \int_0^1 dx \biggl\{
     \frac12 \, \frac{ d\Gamma_\ggg(E,x)/dx }{ \Gamma_{\rm g}(E) }
       \left[
          x \, \ellgE(xE)
        + (1-x) \, \ellgE\bigl((1-x)E\bigr)
       \right]
\\
     + \frac{ d\Gamma_\gqq(E,x)/dx }{ \Gamma_{\rm g}(E) }
       \left[
          x \, \ellqE(xE)
        + (1-x) \, \ellqE\bigl((1-x)E\bigr)
       \right]
  \biggr\} ,
\end {multline}
where
\begin {equation}
   \Gamma_{\rm g} \equiv \Gamma_\ggg + \Gamma_\gqq
   = \frac12 \int_0^{1} dx \> \frac{d\Gamma_\ggg}{dx}
           + \int_0^{1} dx \> \frac{d\Gamma_\gqq}{dx}
\end {equation}
is the total rate for a gluon to split by either bremsstrahlung or
pair production.
We now need the corresponding equation for the quark energy
stopping distance:
\begin {equation}
   \ellqE(E) =
   \frac{1}{\Gamma_{\qgq}(E)}
   + \int_0^1 dx \>
     \frac{ d\Gamma_\qgq(E,x)/dx }{ \Gamma_\qgq(E) }
       \left[
          x \, \ellgE(xE)
        + (1-x) \, \ellqE\bigl((1-x)E\bigr)
       \right] .
\end {equation}

To solve this coupled system of equations, we proceed as before.
First eliminate the issue of infrared divergences by rewriting them
as
\begin {multline}
   \int_0^1 dx \biggl\{
     \frac12 \, \frac{ d\Gamma_\ggg(E,x) }{ dx }
       \left[
          \ellgE(E)
        - x \, \ellgE(xE)
        - (1-x) \, \ellgE\bigl((1-x)E\bigr)
       \right]
\\
     + \frac{ d\Gamma_\gqq(E,x) }{ dx }
       \left[
          \ellgE(E)
        - x \, \ellqE(xE)
        - (1-x) \, \ellqE\bigl((1-x)E\bigr)
       \right]
  \biggr\}
  = 1 ,
\end {multline}
\begin {equation}
   \int_0^1 dx \>
     \frac{ d\Gamma_\qgq(E,x) }{ dx }
       \left[
          \ellqE(E)
        - x \, \ellgE(xE)
        - (1-x) \, \ellqE\bigl((1-x)E\bigr)
       \right] 
   = 1 .
\end {equation}
Then we make the leading-log approximation (\ref{eq:LLscaling}) and
write
\begin {equation}
  \ellgE(E) = \frac{E^{1/2}}{\AgE} \,,
  \qquad
  \ellqE(E) = \frac{E^{1/2}}{\AqE} \,,
\label {eq:lE}
\end {equation}
to
obtain coupled algebraic equations for $1/\AgE$ and
$1/\AqE$.  The solution is
\begin {subequations}
\label {eq:Aenergy}
\begin {equation}
   \AgE = \frac{M_{\rm gg} M_{\rm qq} - M_{\rm gq} M_{\rm qg}}
               {M_{\rm qq} - M_{\rm gq}} ,
   \qquad
   \AqE = \frac{M_{\rm gg} M_{\rm qq} - M_{\rm gq} M_{\rm qg}}
               {M_{\rm gg} - M_{\rm qg}} ,
\end {equation}
where
\begin {align}
   M_{\rm gg} &= 
   \int_0^1 dx \biggl\{
     \frac12 \, \frac{ d\tilde\Gamma_\ggg(x) }{ dx }
       \left[
          1
        - x^{3/2}
        - (1-x)^{3/2}
       \right]
     + \frac{ d\tilde\Gamma_\gqq(E,x) }{ dx } 
   \biggr\} ,
\\
   M_{\rm gq} &= 
   \int_0^1 dx \>
     \frac{ d\tilde\Gamma_\gqq(x) }{ dx }
       \left[
        - x^{3/2}
        - (1-x)^{3/2}
       \right] ,
\\
   M_{\rm qg} &= 
   \int_0^1 dx \>
     \frac{ d\tilde\Gamma_\qgq(x) }{ dx }
       \left[
        - x^{3/2}
       \right] ,
\\
   M_{\rm qq} &= 
   \int_0^1 dx \>
     \frac{ d\tilde\Gamma_\qgq(x) }{ dx }
       \left[
        1
        - (1-x)^{3/2}
       \right] .
\end {align}
\end {subequations}


\subsection {Leading Log Details}
\label {sec:LLdetails}

\subsubsection {Bremsstrahlung and pair production rates}

Following notation similar to Ref.\ \cite{ArnoldXiao}, the
leading-log splitting rates are
\begin {equation}
  \frac{d\Gamma_\abc}{dx} =
  \frac{\alpha \, \mu_\abc^2 \,P_{\rm a{\to}b}(x)}
       {4\pi\sqrt2 \,x(1-x) E}
\label{eq:LLGamma}
\end {equation}
where, at leading-log order,
\begin {multline}
  \mu_\abc^2 \simeq
  \Bigl\{
    4 x(1-x) E
    \bigl[ (-C_a+C_b+C_c)+ (C_a-C_b+C_c)x^2
\\
           + (C_a+C_b-C_c)(1-x)^2 \bigr]
    \qhat(Q_{\perp0})
  \Bigr\}^{1/2} .
\label {eq:LLmu}
\end {multline}
Here $\qhat(\Lambda)$ is
proportional to the average squared transverse momentum $Q_\perp^2$
that a high-energy particle picks up per unit length while traveling
through the plasma.
We'll discuss the formula shortly.
In the present context, the definition of $\qhat(\Lambda)$ is
UV-regulated by imposing an ultra-violet cut-off $\Lambda$ imposed
on the momentum transfers from individual collisions.
$Q_{\perp 0} \sim (\hat q E)^{1/4}$ is any rough estimate
of the
total transverse momentum transfer during one bremsstrahlung formation
time,%
\footnote{
\label{foot:Qperp0}
  Here is a lightning review.
  By the uncertainty principle, the formation time
  $t_{\rm f}$ is of order $1/\Delta E$, where $\Delta E$ is
  the amount by which energy would be violated if an on-shell
  high-energy particle split by
  bremsstrahlung (or pair production) in isolation.  For hard
  bremsstrahlung, $\Delta E \sim Q_\perp^2/E$ in the high-energy limit,
  where $Q_\perp$ is the transverse momentum of the final particles,
  and so $t_{\rm f} \sim E/Q_\perp^2$.  The $Q_\perp$
  must be supplied by collisions with the plasma.
  By definition of $\hat q$, the amount of $Q_\perp$ picked
  up in one formation
  time is given by
  $Q_\perp^2 \sim \hat q t_{\rm f}$.  Putting the last two equations
  together gives $t_{\rm f} \sim (E/\hat q)^{1/2}$ and
  $Q_\perp \sim (\hat q E)^{1/4}$.
  (For a review, see, for example, Sec.\ 3.2 of Ref.\ \cite{BSZ} together with
  Ref.\ \cite{BaierConference}.)
}
and
ambiguities in that guess of $O(1)$ factors only affect
the answer beyond leading-log order.  $C_s$ is the quadratic Casimir
associated with the color representation of particle $s$.  For QCD,
\begin {equation}
   C_{\rm q} = C_{\rm F} = \tfrac43 ,
   \qquad
   C_{\rm g} = C_{\rm A} = 3 .
\end {equation}
The $P_{a{\to}b}$ are the usual Dokshitzer-Gribov-Lipatov-Alterelli-Parisi
(DGLAP) splitting functions
\begin {align}
   P_{{\rm q}\to {\rm g}}(x)
   &= \cf \, \frac{[1+(1-x)^2]}{x} \,,
\\
   P_{{\rm g}\to {\rm g}}(x)
   &= \ca \, \frac{[1 + x^4 + (1-x)^4]}{x(1-x)} \,,
\\
   P_{{\rm g}\to {\rm q}}(x)
   &= \Nf \tf [x^2+(1-x)^2] ,
\label {eq:Pgq}
\end {align}
where $\Nf$ is the number of quark flavors
and $t_s$ is the trace normalization of color generators
$T^a$ defined by
$\tr(T_s^a T_s^b) = t_s \, \delta^{ab}$, with
\begin {equation}
   t_{\rm F} = \tfrac12 \,,
   \qquad
   t_{\rm A} = C_{\rm A} \,,
\end {equation}

The UV-regulated
$\qhat(\Lambda)$ discussed above is given in weak coupling by
\begin {equation}
   \hat q_s(\Lambda) \equiv C_s \qhat(\Lambda)
   = \int_{q_\perp<\Lambda} d^2 q_\perp \>
     \frac{d\Gamma_{{\rm el},s}}{d^2 q_\perp} \, q_\perp^2 ,
\end {equation}
where $\Gamma_{\rm el}$ is the rate of elastic $2{\to}2$
scattering of a high-energy particle of type $s$ off of the plasma,
and $\q_\perp$ is the transverse momentum transfer in an
individual such collision.  The only dependence on the
type of high-energy particle is a factor of $C_s$, which we
have factored out of the definition of $\qhat$.

The weak-coupling result for $\qhat(\Lambda)$ is slightly different depending
on the precise range of $\Lambda$.
At leading-log order in coupling,%
\footnote{
  For a discussion of the perturbative result for $\hat q(\Lambda)$ using
  this notation, see Ref.\ \cite{ArnoldXiao}.
}
\begin {equation}
   \qhat(\Lambda) \simeq
   \alpha T \md^2 \ln\left( \frac{T^2}{\md^2} \right)
    + 4\pi\alpha^2{\cal N} \ln\left( \frac{\Lambda^2}{T^2} \right)
\label {eq:qhatbig}
\end {equation}
if $\Lambda \gtrsim T$, where the second term dominates in the limit of
large $\Lambda$; and
\begin {equation}
   \qhat(\Lambda) \simeq
   \alpha T \md^2 \ln\left( \frac{\Lambda^2}{\md^2} \right)
\label {eq:qhatsmall}
\end {equation}
if $\md \ll \Lambda \lesssim T$.  Here $\md$ is the Debye mass, given by
\begin {equation}
  \md^2 = \tfrac13 ( \ta + \Nf\tf ) g^2 T^2
        = (1 + \tfrac16\,\Nf) g^2 T^2 ,
\end {equation}
and ${\cal N}$ is the plasma particle density weighted by group factors,
given by
\begin {equation}
  {\cal N} = \tfrac13 (\ta + \tfrac32\,\Nf\tf)
             \frac{\zeta(3)}{\zeta(2)} \, T^3
           = (1 + \tfrac14\,\Nf) \frac{\zeta(3)}{\zeta(2)}  \, T^3 ,
\end {equation}
where $\zeta(s)$ is the Riemann zeta function.


\subsubsection {Leading log stopping distances}

Substituting the leading-log bremsstrahlung and pair production rates
into the general leading-log formulas (\ref{eq:LLAq}) and
(\ref{eq:Aenergy}) gives us leading-log stopping distances.
For the quark number stopping distance (\ref{eq:LLAq}),
\begin {multline}
   A_{\rm q}
   = \frac{\alpha}{2\pi\sqrt2}
     [\qhat(Q_{\perp0})]^{1/2}
     \int_0^1 dx \>
        P_{{\rm q}\to {\rm g}}(x)
        \left[
          \frac{ \ca + (2\cf-\ca)x^2 + \ca (1-x)^2 }
               { x(1-x) }
        \right]^{1/2}
\\ \times
     \left[ 1 - (1-x)^{1/2} \right] ,
\label {eq:Aqint}
\end {multline}

Now take
$Q_{\perp0} \sim (\hat q E)^{1/4} \sim (\alpha^2 T^3 E)^{1/4}$.
If $Q_{\perp0} \ll T$, then (\ref{eq:qhatsmall}) gives
\begin {equation}
   \qhat(Q_{\perp0}) \simeq
   \alpha T \md^2 \ln\left( \frac{Q_{\perp0}^2}{\md^2} \right)
   \simeq \tfrac12 \alpha T \md^2
          \ln\left( \frac{\alpha^2 T^3 E}{\md^4} \right)
   \simeq \tfrac12 \alpha T \md^2 \ln\left( \frac{E}{T} \right)
\end {equation}
at leading-log order.  Note that the factors of coupling cancel out
in the logarithm.  If $Q_{\perp 0} \gg T$, and if we keep track
of logarithms of coupling as well as logarithms of energy, then
\begin {align}
   \qhat(Q_{\perp0})
   &\simeq
   \alpha T \md^2 \ln\left( \frac{T^2}{\md^2} \right)
    + 4\pi\alpha^2{\cal N} \ln\left( \frac{Q_{\perp 0}^2}{T^2} \right)
\nonumber\\
   &\simeq
   4\pi\alpha^2 {\cal N} \left[
      \ln\left( \frac{Q_{\perp 0}^2}{\md^2} \right)
      + \frac{c}{2}
        \ln\left( \frac{T^2}{\md^2} \right)
   \right]
\nonumber\\
   &\simeq
   2\pi\alpha^2 {\cal N} \left[
      \ln\left( \frac{E}{T} \right)
      + c \ln\left( \frac{1}{\alpha} \right)
   \right]
\nonumber\\
   &\simeq
   2\pi\alpha^2 {\cal N}
      \ln\left( \frac{E}{\alpha^c T} \right)
\end {align}
with
\begin {equation}
        c = 2 \left[ \frac{T \md^2}{4\pi\alpha{\cal N}} - 1 \right].
\end {equation}
In what follows, it will be convenient to define the dimensionless
numbers
\begin {align}
   \hatmd^2 &\equiv \frac{\md^2}{\alpha T^2}
   = \frac{4\pi}{3} ( \ta + \Nf\tf ) ,
\\
   \hatN &\equiv \frac{4\pi{\cal N}}{T^3}
   = \frac{4\pi}{3} (\ta + \tfrac32\,\Nf\tf)
             \frac{\zeta(3)}{\zeta(2)} \,.
\end {align}

We can now summarize the leading-log result as being given by
(\ref{eq:qstop}) with
\begin {equation}
   a
   = \frac{\hat{\cal Q}^{1/2}}{4 \pi}
     \int_0^1 dx \>
        P_{{\rm q}\to {\rm g}}(x)
        \left[
          \frac{ \ca + (2\cf-\ca)x^2 + \ca (1-x)^2 }
               { x(1-x) }
        \right]^{1/2}
     \left[ 1 - (1-x)^{1/2} \right] ,
\label {eq:a}
\end {equation}
\begin {equation}
   \hat{\cal Q} \equiv
   \begin {cases}
      {\hatN} ,
        & Q_\perp \gg T; \\
      \hatmd^2 ,
        & Q_\perp \ll T ,
   \end {cases}
\end {equation}
and
\begin {equation}
  c =
  \begin {cases}
    2 \left( \frac{\hatmd^2}{\hatN} - 1 \right) ,
      & Q_\perp \gg T; \\
    0 ,
      & Q_\perp \ll T .
   \end {cases}
\label {eq:c}
\end {equation}
Plugging in QCD group factors and doing the integral (\ref{eq:a})
numerically gives the results for $a$ and $c$ in Tables
\ref{tab:abc} and \ref{tab:abcsmall} with
\begin {equation}
   a = 0.55634 \, {\hat{\cal Q}}^{1/2}
\end {equation}
The result listed for one-flavor QED corresponds to setting $\ca=\ta=0$
and $\cf=\tf=1$.

A similar analysis of the energy stopping distances of (\ref{eq:lE}) gives
\begin {equation}
  \ell^{\rm(e)}_{s}
  \simeq \frac{1}{a^{\rm(e)}_{s} \alphas^2 T} \sqrt{ \frac{E}{T L} }
\end {equation}
where, at leading-log order,
\begin {equation}
  L \simeq \ln\left(\frac{E}{\alphas^c T}\right)
\end {equation}
as before, with the same value (\ref{eq:c}) of $c$.
The
coefficients $a^{\rm(e)}_s$, however, are given by
\begin {equation}
   a^{\rm(e)}_{\rm g} =
     \frac{\hat{\cal Q}^{1/2}}{4 \pi}
     \frac{(m_{\rm gg} m_{\rm qq} - m_{\rm gq} m_{\rm qg})}
          {m_{\rm qq} - m_{\rm gq}} ,
   \qquad
   a^{\rm(e)}_{\rm q} =
     \frac{\hat{\cal Q}^{1/2}}{4 \pi}
     \frac{(m_{\rm gg} m_{\rm qq} - m_{\rm gq} m_{\rm qg})}
          {m_{\rm gg} - m_{\rm qg}} ,
\end {equation}
where
\begin {align}
   m_{\rm gg}
   &=
     \int_0^1 dx \> \biggl\{
        \tfrac12 \, P_{{\rm g}\to {\rm g}}(x)
        \left[
          \frac{ \ca + \ca x^2 + \ca (1-x)^2 }
               { x(1-x) }
        \right]^{1/2}
        \left[ 1 - x^{3/2} - (1-x)^{3/2} \right]
\nonumber\\ & \qquad\qquad
      + P_{{\rm g}\to {\rm q}}(x)
        \left[
          \frac{ (2\cf-\ca) + \ca x^2 + \ca (1-x)^2 }
               { x(1-x) }
        \right]^{1/2}
     \biggr\} ,
\\
   m_{\rm gq}
   &=
     \int_0^1 dx \>
         P_{{\rm g}\to {\rm q}}(x)
        \left[
          \frac{ (2\cf-\ca) + \ca x^2 + \ca (1-x)^2 }
               { x(1-x) }
        \right]^{1/2}
     \left[ -x^{3/2}-(1-x)^{3/2} \right] ,
\\
   m_{\rm qg}
   &=
     \int_0^1 dx \>
         P_{{\rm q}\to {\rm g}}(x)
        \left[
          \frac{ \ca + (2\cf-\ca) x^2 + \ca (1-x)^2 }
               { x(1-x) }
        \right]^{1/2}
     \left[ -x^{3/2} \right] ,
\\
   m_{\rm qq}
   &=
     \int_0^1 dx \>
         P_{{\rm q}\to {\rm g}}(x)
        \left[
          \frac{ \ca + (2\cf-\ca) x^2 + \ca (1-x)^2 }
               { x(1-x) }
        \right]^{1/2}
     \left[ 1 - (1-x)^{3/2} \right] .
\end {align}
Results are displayed in Tables \ref{tab:abc} and \ref{tab:abcsmall}.


\subsection {Quark mass thresholds}

In this paper, $\Nf$ is the number of effectively massless quark
species.  Here we comment on what counts as a massless quark for the
purposes of our analysis.  Parametrically, the condition to ignore
the mass of a high-energy quark in a splitting process $q \to q g$
or $g \to q\bar q$ is%
\footnote{
  See footnote \ref{foot:Qperp0}.  For hard bremsstrahlung,
  the quark mass will give a
  contribution of order $M^2/E$ to $\Delta E \sim Q_\perp^2/E$.
  This effect of the mass is negligible when $M \ll Q_\perp$.
  Plugging in the massless result $Q_\perp \sim (\hat q E)^{1/4}$
  makes this condition (\ref{eq:M}).
}
\begin {equation}
   M \ll Q_\perp \sim (\hat q E)^{1/4} .
\label {eq:M}
\end {equation}
So we can use our formulas to describe the quark number stopping time
of such a quark, or we can compute energy stopping times provided
we take $\Nf$ in the DGLAP splitting function (\ref{eq:Pgq}) to
be the number of quark species that satisfy (\ref{eq:M}).
All other factors of $\Nf$ in this paper refer to the number
of quark species in the plasma, which should be taken as
the number satisfying $M \ll T$.
The specific results in tables
\ref{tab:abc} and \ref{tab:abcsmall} are given for the case that these
two $\Nf$ values are the same.


\section{NLLO Analysis of Quark Number Stopping}
\label {sec:NLLO}

Now we extend the analysis of the previous section to
next-to-leading logarithmic order for the case of the quark number
stopping distance.  Our goal is to determine the coefficient $b$
in (\ref{eq:qstop}).


\subsection {NLLO splitting rates}

The NLLO result for splitting rates was computed in
Refs.\ \cite{ArnoldDogan,ArnoldXiao} and corresponds to
the leading-log result (\ref{eq:LLGamma}),
\begin {subequations}
\label {eq:G}
\begin {equation}
  \frac{d\Gamma_\abc}{dx} =
  \frac{\alpha \, \mu_\abc^2 \,P_{\rm a{\to}b}(x)}
       {4\pi\sqrt2 \,x(1-x) E} \,,
\label{eq:Gamma}
\end {equation}
with the
formula (\ref{eq:LLmu}) for $\mu_\abc^2$ replaced by
\begin {multline}
  \mu_\abc^2 \simeq
  \biggl\{
    4 x(1-x) E
    \biggl[ (-C_a+C_b+C_c) \,
               \qhat\left( \xi^{1/2}\mu_\abc \right)
         +  (C_a-C_b+C_c)x^2 \,
               \qhat\left( \frac{\xi^{1/2}\mu_\abc}{x} \right)
\\
           + (C_a+C_b-C_c)(1-x)^2 \,
               \qhat\left( \frac{\xi^{1/2}\mu_\abc}{1-x} \right)
    \biggr]
  \biggr\}^{1/2} ,
\label {eq:mu}
\end {multline}
\end {subequations}
where
\begin {equation}
   \xi \equiv \exp( 2 - \gammaE + \tfrac{\pi}4 ) .
\end {equation}
Numerically, equation (\ref{eq:mu})
may be solved self-consistently for $\mu_\abc$.
Alternatively, one may solve it iteratively starting from an initial
guess $\mu_{\abc} = Q_{\perp0} \sim (\hat q E)^{1/4}$
on the right-hand side, then
using the result as a refined guess, and so forth.  This generates
an explicit expansion of the solution in inverse powers of logarithms.
In either case, the final result for the rate $d\Gamma/dx$ is only
correct to NLLO.  For the quark-number stopping distance, we only
care about the $q\to gq$ splitting rate.

The explicit expansion of the
rate (\ref{eq:Gamma}) in powers of inverse logarithms has the form
\begin {equation}
  \frac{d\Gamma}{dx} =
  \alpha^2 T \left( \frac{T}{E} \right)^{1/2}
  {\cal A}(x) \, \ln^{1/2}(\kappa E) \left[1 +
     \frac{ \tfrac12 \ln[\ln(\kappa E)]
            + {\cal B}(x) }
          {\ln(\kappa E)}
     + \cdots
  \right] ,
\label {eq:Gform}
\end {equation}
where $\kappa$ is an arbitrary constant with dimensions of inverse
energy.  ${\cal B}(x)$ and higher-order coefficients in the
expansion implicitly depend on $\kappa$ in such a way that the
rate is formally independent of the choice of $\kappa$ --- that is,
the dependence on $\kappa$ at any fixed order in the expansion is
always a higher-order effect.  However, in practice $\kappa$ should
be chosen to be parametrically of order $1/\alpha^c T$.  (We will
make a much more specific choice of $\kappa$ later on.)

${\cal A}(x)$ simply gives the leading-log term, which we will
find notationally convenient to write as (specializing to
$q\to gq$)
\begin {equation}
  {\cal A}(x) =
  \frac{[\lambda(x)]^{1/2}}{4\pi x(1-x)} \,
  P_{{\rm q}{\to}{\rm g}}(x)
\end {equation}
with
\begin {equation}
  \lambda(x) \equiv
  \hat{\cal Q} x(1-x)
    \left[ \ca + (2\cf - \ca)x^2 + \ca(1-x)^2 \right]
  .
\end {equation}

For NLLO, we need better than the leading-log result
(\ref{eq:qhatbig}) for $\qhat(\Lambda)$.
Complete results to leading order in powers of coupling are presented in
Refs.\ \cite{ArnoldXiao,Simon}.%
\footnote{
  Ref.\ \cite{Simon} also calculated the next-order correction in powers
  of the coupling $g$, which is quite significant for any value of
  $\alphas$ of possible interest to experiment.
}
For the case $\Lambda \gg T$, 
\begin {equation}
   \qhat(\Lambda) \simeq
   \alpha^2 T^3
   \left[
     \hatmd^2 \ln\frac{\xi' T^2}{\md^2}
     + \hatN \ln\frac{\Lambda^2}{\xi' T^2}
     - \tfrac{16}{\pi} ( \ta\sigma_+ + 2\Nf\tf\sigma_- )
   \right]
\label {eq:qhatNLLO}
\end {equation}
where
\begin {equation}
  \xi' \equiv 4 \exp\left( 1 - 2\gammaE \right) ,
\end {equation}
and
\begin {align}
  \sigma_+ &\equiv \sum_{k=1}^\infty \frac{1}{k^3} \, \ln[(k-1)!]
           \simeq 0.3860438 \,,
\\
  \sigma_- &\equiv \sum_{k=1}^\infty \frac{(-)^{k-1}}{k^3} \, \ln[(k-1)!]
           \simeq 0.0112168 \,.
\end {align}
For the case $\Lambda \ll T$, the leading-order answer for $\qhat$ is
in fact the same as (\ref{eq:qhatsmall})
\begin {equation}
   \qhat(\Lambda) \simeq
   \alpha T \md^2 \ln\left( \frac{\Lambda^2}{\md^2} \right) .
\end {equation}

Combining these expressions for $\qhat(\Lambda)$ with (\ref{eq:G})
and (\ref{eq:Gform}) gives
\begin {equation}
  {\cal B}(x) =
  \tfrac12\ln\left( \frac{1}{\alpha^c T \kappa} \right)
  + {\cal B}_1(x) + {\cal B}_2(x),
\end {equation}
with
\begin {subequations}
\label {eq:B12}
\begin {equation}
  {\cal B}_1(x) = 
  \tfrac12 \ln[2 \, \lambda(x)]
  - 2 \,
    \frac{
      \left[ (2\cf - \ca)x^2 \ln x + \ca(1-x)^2 \ln(1-x) \right]
    }{
      \left[ \ca + (2\cf - \ca)x^2 + \ca(1-x)^2 \right]
    } \,,
\end {equation}
\begin {equation}
  {\cal B}_2(x) =
  \begin {cases}
    \ln\frac{\xi}{\xi'} +
    \frac{1}{\hatN} \left[
       \hatmd^2\ln\frac{\xi'}{\hatmd^2}
       - \frac{16}{\pi}(\ta\sigma_++2\Nf\tf\sigma_-)
    \right]
    ,  & Q_\perp \gg T ; \\
    \ln\frac{\xi}{\hatmd^2}
    , & Q_\perp \ll T .
  \end {cases}
\end {equation}
\end {subequations}


\subsection {NLLO stopping distance}

Corresponding to the expansion (\ref{eq:Gform}) of the rate, one
may now look for a solution to the quark number stopping distance
equation (\ref{eq:tq2}) in the form
\begin {equation}
   \ell_{\rm q}(E)
   = \frac{E^{1/2}}{A \ln^{1/2}(\kappa E)} \left[1 -
     \frac{ \tfrac12 \ln[\ln(\kappa E)]
            + B }
          {\ln(\kappa E)}
     + \cdots
  \right] .
\label {eq:tansatz}
\end {equation}
The internal minus sign in this formula compared to (\ref{eq:Gform}) arises
because the length is related to the inverse of the rate.
Plugging the ansatz (\ref{eq:tansatz}) and the NLLO rate (\ref{eq:Gform})
into (\ref{eq:tq2}) determines the same
$A = a \alpha^2 T^{3/2}$ as in the leading-log calculation, with
\begin {equation}
  a = \int_0^1 dx \> {\cal A}(x) \, \left[1-(1-x)^{1/2}\right] .
\end {equation}
It also determines
\begin {align}
  B &=
  \frac{1}{a} \int_0^1 dx \> {\cal A}(x) \, \left\{
   {\cal B}(x) \left[1-(1-x)^{1/2}\right]
   + \tfrac12 (1-x)^{1/2}\ln(1-x)
 \right\}
\nonumber\\
  &=
  - \tfrac12 \ln(\alpha^c T \kappa) + \tfrac12 \beta
\end {align}
with
\begin {equation}
  \beta \equiv
   \frac{1}{a} \int_0^1 dx \> {\cal A}(x) \, \left\{
   2[{\cal B}_1(x)+{\cal B}_2(x)] \left[1-(1-x)^{1/2}\right]
   + (1-x)^{1/2}\ln(1-x)
 \right\} .
\label {eq:beta}
\end {equation}

As mentioned earlier, the precise choice of $\kappa$ is formally arbitrary,
and the effect of changing $\kappa$ by a multiplicative factor of
order one does not affect the NLLO result except by corrections that are yet
higher order in inverse logarithms.  However, as a practical matter,
it is useful to have some definite prescription for choosing a sensible
value for $\kappa$.  We will use the fastest apparent convergence (FAC)
prescription, which in this context is to choose $\kappa$ such
that the NLLO correction in (\ref{eq:tansatz}) vanishes:
\begin {equation}
   \tfrac12 \ln[\ln(\kappa E)]
   - \tfrac12 \ln(\alpha^c T \kappa) + \tfrac12 \beta = 0 ,
\end {equation}
and so
\begin {equation}
  \ln(\kappa E) =
  \ln\left( \frac{e^{\beta} E \ln(\kappa E)}{\alpha^c T} \right) .
\end {equation}
This corresponds to (\ref{eq:L}) with the identification
$L = \ln(\kappa E)$ and
\begin {equation}
  b = e^{\beta} .
\label {eq:bbeta}
\end {equation}
The combination of
equations (\ref{eq:B12}), (\ref{eq:beta}), and (\ref{eq:bbeta}) is
our final result for the NLLO coefficient $b$ in (\ref{eq:qstop}).
Evaluating the integrals numerically gives the tabulated results
for $b$ in Tables \ref{tab:abc} and \ref{tab:abcsmall}.


\section {A numerical test of the result}
\label {sec:numerics}

In this paper, we have derived a simple NLLO formula
(\ref{eq:qstop}) for the quark number stopping distance of
a high-energy particle.
This result relies on an expansion in powers of inverse logarithms,
which parametrically requires $\ln(E/T) \gg 1$, and one may wonder
just how large $E$ has to be before this assumption becomes
reasonable.  As an alternative, one could imagine dropping the
assumption that logarithms are large and instead computing the
stopping distance using the full weak-coupling result for the bremsstrahlung
rate, at leading order in $\alphas$
\cite{AMYsansra,AMYkinetic,AMYx,JeonMoore}.
The disadvantage is that one must then do (somewhat complicated)
numerics---we know of no simple
formula for the stopping distance unless one assumes $\ln(E/T) \gg 1$.
In this section, we perform such a numerical analysis simply as
a check that our result (\ref{eq:qstop}) is indeed correct at
sufficiently large $E$.

The numerics consist of Monte Carlo evolution of a large
sample of quarks with energy $E$, using the full weak-coupling
result%
\footnote{
  Specifically, we take $d\Gamma_\abc/dx = (2\pi)^3 \gamma_\abc/E\nu_a$
  where the $\gamma_\abc$ are the splitting functions of
  Refs.\ \cite{AMYkinetic,AMYx} and the number $\nu_a$ of spin+color degrees of
  freedom is 6 for a quark or anti-quark and 16
  for a gluon.  We do not include any final-state Bose enhancement or
  Fermi blocking factors for the final-state particles $b$ and $c$:
  In the limit that their energies are large compared to $T$, no such
  factors are necessary.
}
for the bremsstrahlung rate $d\Gamma/dx$
to randomly determine whether each quark loses energy $xE$ 
in each small time step $\Delta t$.  We will focus on the
case $E \gg T$ and so may (for simplicity)
ignore other mechanisms of energy loss,
such as collisional energy loss.
For the same reason, we only consider energy loss and will ignore
processes which can increase the quark energy (such as inverse
bremsstrahlung).
Also for the sake of simplicity, we will formally
focus on the $Q_\perp \ll T$
case (parametrically $E \ll T/\alphas^2\ln(\alphas^{-1})$), which is
the approximation implicitly used in previous numerical calculations
of the bremsstrahlung rate for weak coupling.%
\footnote{
  Specifically, numerical calculations such as Refs. \cite{AMYx,JeonMoore}
  have used the formula
  ${\cal A}(q_\perp) = \md^2 T/q_\perp^2(q_\perp^2+\md^2)$ in the
  notation of Ref.\ \cite{AMYx}.
  (${\cal A}(q_\perp)$ corresponds to
  $(2\pi/g)^2 d\Gamma_{\rm el}/{d^2q_\perp}$ in this paper's
  notation and
  $T\,C(q_\perp)$ in the notation of Ref.\ \cite{JeonMoore}.)
  This formula is correct only when $Q_\perp \ll T$.
  As noted in Ref.\ \cite{ArnoldDogan}, for 3-flavor QCD, use of
  this formula in the $Q_\perp \gg T$ case should produce
  an error of at most 15\%.
}
In the numerics, we will call a quark ``stopped''
when its energy drops below $T$, since our approximations no longer
make sense then and since a real quark would equilibrate once its
energy dropped to be of order $T$.  Whether one defines ``stopped''
as $E< T$ or $E< T/2$ or $E < 2T$ will only affect the stopping
distance by a relative correction that is parametrically of order
$\sqrt{T/E}$ (up to logarithms) and so is unimportant when
$E \gg T$.

In each time step, we let each quark bremsstrahlung with probability
$\Gamma \, \Delta t$.  When bremsstrahlung occurs, we
randomly choose the momentum fraction
of the bremsstrahlung gluon according to the probability distribution
$\Gamma^{-1} \, d\Gamma/dx$.  This is not completely straightforward
because the total bremsstrahlung rate $\Gamma$ has a small $x$
divergence.  Radiation of small $x$ gluons is inefficient for energy
loss, and so the final result for the stopping distance should not
be sensitive to this divergence.  In our numerics, we simply sidestep
such divergences by restricting the range of $x$ values to
$\epsilon_1 \le x \le 1-\epsilon_2$, where $\epsilon_1$ and
$\epsilon_2$ are small, and then we check that our numerical results
converge as we make $\epsilon_1$ and $\epsilon_2$ (and $\Delta t$)
smaller and smaller.

\begin {figure}
\begin {center}
  \includegraphics[scale=1.0]{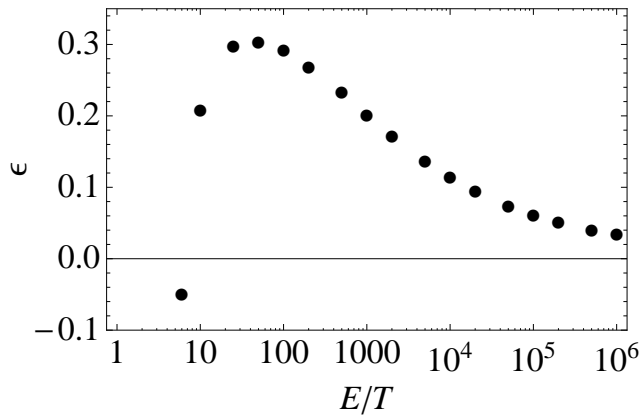}
  \caption{
     \label{fig:relerr}
     The relative difference $\epsilon \equiv 1-t_{\rm NLLO}/t_{\rm numeric}$
     between (i) the NLLO quark number stopping distance (\ref{eq:qstop}) and
     (ii) the quark number stopping distance $t_{\rm numeric}$ computed
     through simulation
     using the full weak-coupling result for the bremsstrahlung rate.
     The slight scatter of points is due to the statistics of simulating
     only a finite number of particles (in this case, $10^5$ particles).
  }
\end {center}
\end {figure}

A comparison of the full numeric result with the NLLO approximation
(\ref{eq:qstop}) is shown in Fig.\ \ref{fig:relerr} as a function
of $E/T$ for 3-flavor QCD.  The plot shows the size of the
relative error $\epsilon \equiv 1-t_{\rm NLLO}/t_{\rm numeric}$ that the
NLLO approximation makes compared to the numeric simulation.
The plot extends to absurdly large values of $E/T$ because
the purpose of this plot is to check our NLLO analysis.  We can see
that the NLLO result indeed approaches the full weak-coupling result
at very large $E$.  Moreover, it does not do too badly even at smaller
values of $E/T$ that are only moderately large: The difference
is $\lesssim 30\%$ for $E \gtrsim 6 T$.  One should not
take this too seriously, however, since at $E \sim 6 T$ the various
approximations we made earlier in this section are in doubt.  We do not
show any comparison for even smaller $E/T$ simply because the NLLO result
(\ref{eq:qstop}), which implements the FAC prescription, ceases
to exist: There is no solution to (\ref{eq:L}) for $L$ when
$E/T \le \alphas^c e/b$, which in the present case is
$E/T \le 5.74$.  At exactly $E/T = \alphas^c e/b$, the logarithm $L$
becomes $1$, and so the idea of a large-logarithm approximation has
certainly broken down then.

\begin {figure}
\begin {center}
  \includegraphics[scale=1.0]{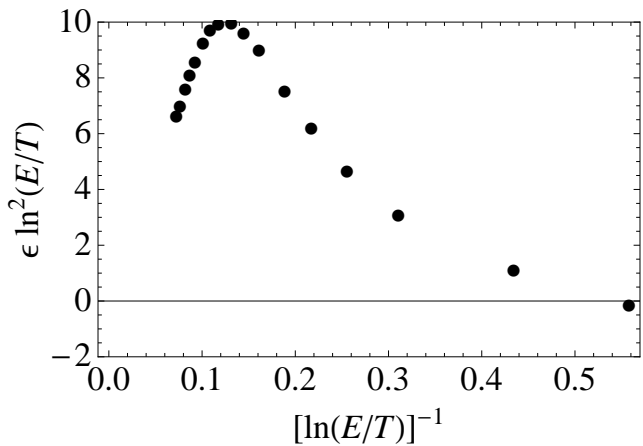}
  \caption{
     \label{fig:check}
     The results of Fig.\ \ref{fig:relerr} plotted with axes
     convenient for checking the correctness of the NLLO result
     at very large $E$.
  }
\end {center}
\end {figure}

As a further check of our NLLO calculation, we plot
$\epsilon \ln^2(E/T)$ versus $y \equiv 1/\ln(E/T)$ in
Fig.\ \ref{fig:check}.  If the NNLO contribution was incorrect,
the result would diverge as $y \to 0$.

\begin {figure}
\begin {center}
  \includegraphics[scale=0.3]{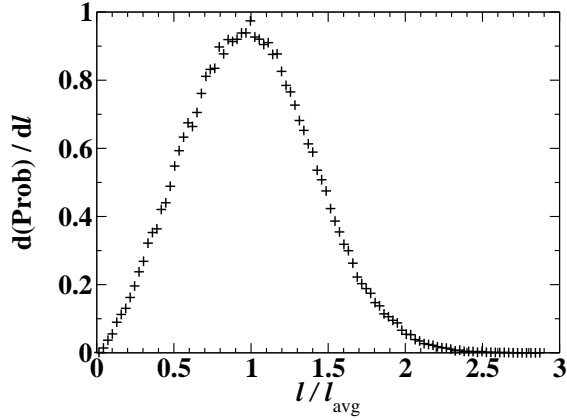}
  \caption{
     \label{fig:distribution}
     Simulation results for the
     probability distribution for quarks to lose their energy
     through bremsstrahlung in a distance $\ell$, as function of
     $\ell/\ell_{\rm avg}$, where $\ell_{\rm avg}$ is the average.
     The particular results shown are for $E/T = 10^6$, for which
     $\ell_{\rm avg}$ agrees with the NLLO stopping length
     (\ref{eq:qstop}) to within 4\%.
  }
\end {center}
\end {figure}

Finally, just to give some idea of the distribution of stopping lengths
in the high-energy limit, we plot the distribution from our
highest-energy numerical
simulation in Fig.\ \ref{fig:distribution}.
For some quantities of interest in the subject of energy loss (such
as the medium effect on energy loss from passage through a
thin medium), there
are slowly falling tails to distributions which cause average values
of those quantities to be unrepresentative of typical events.  That is not
the case here for the stopping length:
the average coincides with the location of the peak of the distribution.

\section{Running Coupling}
\label {sec:running}

Our result (\ref{eq:qstop}) formally assumes that $\alpha \ln(E/T) \ll 1$
and so ignores, for instance, running of the coupling $\alpha$.
One can plausibly accommodate running in the bremsstrahlung rate, and so
the stopping distance, by the prescriptions reviewed in 
Ref.\ \cite{ArnoldXiao}.%
\footnote{
  See also the earlier discussion in Refs.\
  \cite{ArnoldDogan,Peshier,timelpm1,BDMPS3}.
}
The $\alpha^2$ in the denominator of (\ref{eq:qstop}) comes from
(i) the overall factor of $\alpha$ in (\ref{eq:LLGamma}) associated with
the emission vertex for the bremsstrahlung gluon, and (ii)
the factor $\mu^2$ in (\ref{eq:LLGamma}), which is proportional to
$\sqrt{\hat q} \propto \sqrt{\alpha^2}$.  As reviewed in
Ref.\ \cite{ArnoldXiao}, the emission vertex factor should plausibly
be evaluated at the scale $Q_\perp \sim \mu \sim (\hat q E)^{1/4}$,
and the running of $\alpha$ in $\hat q$ replaces%
\footnote{
  This is a slight simplification.  This nice compact formulation of
  the prescription only
  works if there are no quarks masses between $\md$ and $\Lambda$.
}
\begin {equation}
  \hat q(\Lambda) \propto \alpha^2
  \quad \longrightarrow \quad
  \hat q(\Lambda) \propto \alpha(\Lambda) \, \alpha(\md) .
\label {eq:running}
\end {equation}
Since we needed $\hat q$ at $\Lambda \sim Q_\perp$, the effect of
both these replacements is to replace (\ref{eq:qstop}) by
\begin {equation}
  \ell_{\rm stop,q} \longrightarrow
  \frac{1}{a \, \alpha^{3/2}(Q_\perp) \, \alpha^{1/2}(\md) \, T}
  \sqrt{ \frac{E}{TL} }
\end {equation}
with $Q_\perp \sim (\hat q E)^{1/4}$.  The origin of the $\alpha^c$
in (\ref{eq:L}) is the difference between the scales $\md$ and $T$
in the arguments of the logarithms of (\ref{eq:qhatNLLO}), and so
that $\alpha^c$ should simply be $[\alpha(\md)]^c$.  Throughout this
discussion, we do not distinguish between $\alpha(T)$ and
$\alpha(\md)$ since their difference is small when $\alpha$ is small.

The result (\ref{eq:running}) only accounts for
$\alpha \ln(E/T)$ corrections associated with running
of the coupling.  We are not sure whether there might be other
corrections of comparable size, which is why our introduction is
only so bold as to claim a definite result in the formal limit
$\alpha \ln(E/T) \ll 1$.


\begin{acknowledgments}

We are indebted to Paul Chesler, Larry Yaffe, and Guy Moore
for useful discussions.
This work was supported, in part, by the U.S. Department
of Energy under Grant No.~DE-FG02-97ER41027.

\end{acknowledgments}



\end{document}